\documentclass[useAMS,usenatbib,referee]{mn2e}
\usepackage{epsfig}
\usepackage{graphicx}
\usepackage{lscape}
\usepackage{color}

\title[GeV-quiet soft gamma-ray pulsars]
{Emission mechanism of {\it GeV-quiet} soft gamma-ray pulsars; A case 
for peculiar geometry?}
\author[Wang et al.]{Y. Wang, C.W. Ng, 
J. Takata, 
 Gene C.K. Leung  and  K.S. Cheng
\\
Department of Physics, University of Hong Kong,
Pokfulam Road, Hong Kong;rubyngcw@connect.hku.hk, takata@hku.hk
} 
\begin{document}

\date{}

\pagerange{\pageref{firstpage}--\pageref{lastpage}} \pubyear{2010}

\maketitle

\label{firstpage}

\begin{abstract}
There is a growing new class of young spin-down powered pulsars 
called GeV-quiet soft gamma-ray pulsar; (1) spectral turnover appears  
around~10MeV, (2)  the  X-ray spectra of below 20 keV 
can be described by power law with photon index around 1.2 and (3)  
 the light  curve in X-ray/soft gamma-ray bands shows single broad pulse. 
 Their emission properties are distinct from the normal gamma-ray 
 pulsars, for which the spectral peak in $\nu F_{\nu}$ 
appears in GeV energy bands and the X-ray/gamma-ray light curves 
show sharp and double (or more)  peaks.  In this paper, 
we discuss that X-ray/soft gamma-ray emissions 
of the GeV-quiet soft gamma-ray pulsars are caused by
 the synchrotron radiation of the electron/positron pairs, 
which are created by the magnetic pair-creation process near the 
stellar surface.   In our model,  
the viewing geometry is crucial factor to discriminate between the normal 
gamma-ray pulsars and soft gamma-ray pulsars.  Our  model suggests that 
the difference between the magnetic inclination angle ($\alpha$) 
and the Earth viewing angle ($\beta$) of the soft gamma-ray pulsars 
is small, so that the synchrotron 
emissions from the high magnetic field region around the polar cap region
 dominates in the observed emissions. Furthermore, 
the inclination angle of the soft gamma-ray pulsar  is relatively 
small, $\alpha\leq 30$~degree, and our line of sight is  out of the 
gamma-ray beam emitted via the curvature radiation process in 
 the outer gap. We also 
analysis the six year $Fermi$ data for four soft gamma-ray pulsars to determine
 the  upper limit of the GeV flux.

\end{abstract}


\section{Introduction}
\label{intro}
The $Fermi$ gamma-ray telescope has discovered about 
150 $\gamma$-ray pulsars\footnote{For updated list, 
see https://confluence.slac.stanford.edu/display/GLAMCOG/Public+List+of+LAT-Detected+Gamma-Ray+Pulsars}. The $Fermi$ revealed that the pulsars with  high-spin 
down power emit the GeV gamma-rays and 
the typical gamma-ray spectra are described 
by the single power law plus exponential cut-off function with 
a cut-off energy $\sim$GeV.  It is now widely accepted that 
the GeV gamma-ray emission region locates in 
outer magnetosphere near the light cylinder, 
where the co-rotation speed with the pulsar becomes 
the speed of light (Aliu et al. 2008; Abdo et al. 2010b). 

\begin{figure}
\centering
\includegraphics[width=1.\textwidth]{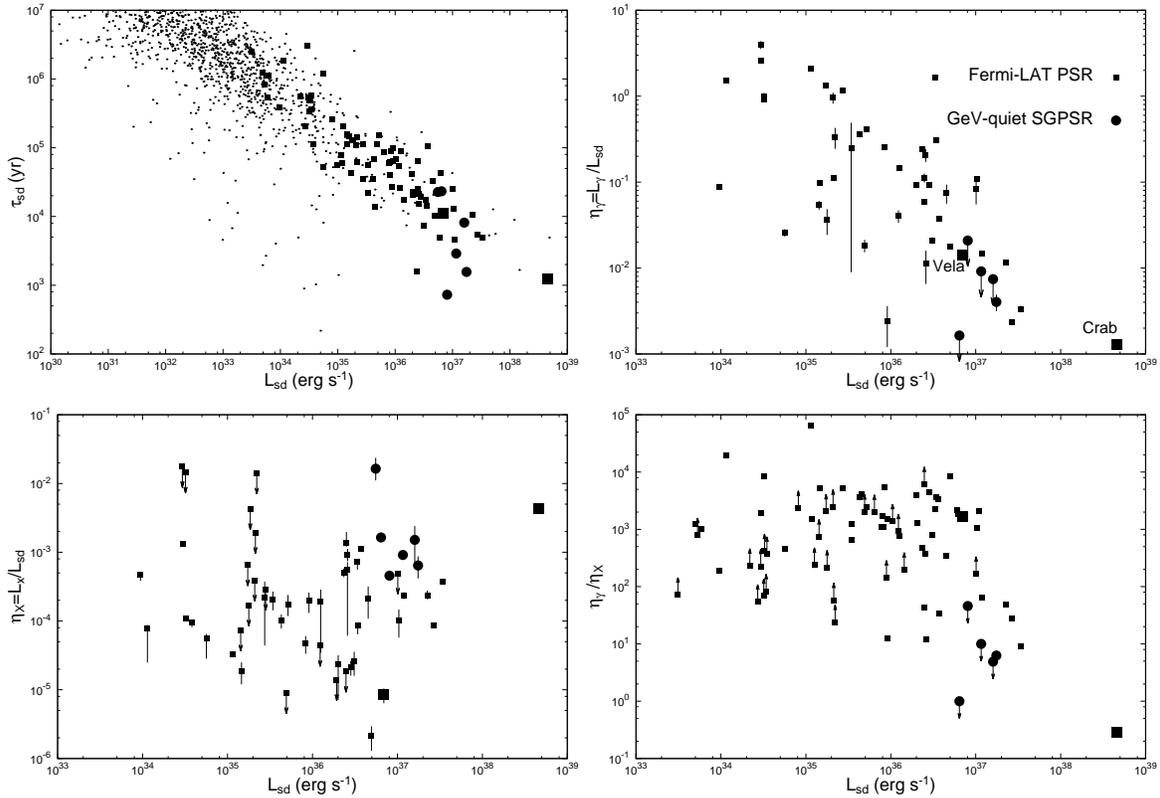}
\caption{Characteristics of the canonical pulsars. Small dots, filled boxes 
and filled circles show the radio pulsars, {\it Fermi}-LAT pulsars and 
GeV-quiet soft gamma-ray pulsars, respectively. The large filed boxes located 
at $L_{sd}=7\times 10^{36}{\rm erg s^{-1}}$ and $4.6\times 10^{38}{\rm erg s^{-1}}$ correspond to the Vela and Crab pulsars, respectively. 
Top left: Spin down power v.s. the characteristic age. Top right: Spin down 
power v.s. gamma-ray efficiency. $L_{\gamma}=4\pi D^2F_{\gamma}$, where $D$ is the distance and $F_{\gamma}$ is the observed flux above 100MeV. Bottom left: 
Spin down power v.s. X-ray efficiency. $L_{X}=4\pi D^2F_{X}$, where 
 $F_{X}$ is the observed X-ray flux below  10keV. Bottom right: Spin down 
power v.s. $\eta_{\gamma}/\eta_{X}$. We extensively used the 
ATNF pulsar catalog  (Manchester et al. 2005) and the $Fermi$ second 
catalog of the pulsars (Abdo et al. 2013). We referred  the observed 
 X-ray flux of GeV-quiet soft gamma-ray pulsars from 
Becker \& Aschenbach (20002) for J1617-5055,
 Torii et al. (1997) for J1811-1925, Lin et al. (2009) for J1838-0655, 
Gotthelf et al. (2000) for J1846-0258 and  Camilo et al. (2002) 
for J1930+1852, respectively.}
\label{sgr}
\end{figure}

The {\it soft gamma-ray pulsar} is growing 
new class of young spin-down powered pulsars that are observed in the 
 non-thermal X-rays and soft gamma-ray bands (Kuiper \& Hermsen, 2013, 2014). 
These soft gamma-ray pulsars are divided into two groups, that is, 
 GeV-loud (e.g. Crab and Vela pulsars) and GeV-quiet. Currently, 
six GeV-quiet soft gamma-ray pulsars (hereafter, GeV-quiet SGPSRs) 
have been known; PSRs B1509-58, J1617-5055, J1811-1925, J1838-0655, 
J1846-02658 and J1930+1852. Figure~\ref{sgr} summarizes the spin down power, 
characteristic age  and spectral characteristics 
of the radio pulsars (small-dots), $Fermi$-LAT pulsars (filled-boxes)
 and GeV-quiet SGPSRs (filled-circles). We can see in Figure~\ref{sgr} that 
GeV-quiet SGPSRs have a relatively large spin down power and 
small characteristics age. Furthermore, we find in Figure~\ref{sgr}  that  
the spectral properties of  GeV-quiet SGPSRs 
 are distinct  from those of  the $Fermi$-LAT  pulsars, 
that is, the weaker gamma-ray emissions but a stronger X-ray emissions comparing with the $Fermi$-LAT  pulsars. In fact, all of GeV-quiet SGPSRs show 
(1) no GeV emissions and  (2)  a single broad light curves 
in X-ray/soft gamma-ray bands.  
The original one, PSR B1509-58,  
was firstly  recognized as the Crab-type pulsar (Ulmer et al. 1993), since 
the spectral peak appears in  $\sim 1$MeV energy, which is resemble 
to the spectrum of the Crab pulsar (Kuiper et al. 2001). 
Unlike the Crab pulsar, however, the off-set of radio/X-ray  
peak phases is fairly large (Abdo et al. 2010a). 
Moreover, the  $Fermi$ revealed that  PSR B1509-58 is not bright in 
GeV gamma-ray bands (Abdo et al. 2010a), which is incompatible with 
 the spectrum of the Crab pulsar, 
suggesting the X-ray/gamma-ray emission mechanism of the PSR B1509-58 
is different from that of the Crab pulsar. In addition to PSR B1509-58, 
PSRs  J1617-5055 (Torii et al. 1998), J1811-1925 (Torii et al. 1997),
 J1838-0655 (Lin et al. 2009),  J1846-0258 (Gotthelf et al. 2000) and  
J1930+1852 (Camilo et al. 2002) are 
classified as GeV-quiet soft gamma-ray pulsars. 
Except for PSR J1617-5055, all of the soft gamma-ray 
pulsars are found in the center of supernova remnants.

The formation of GeV-quiet soft gamma-ray spectra has not been conclusive yet.
Zhang \& Cheng (2000) discussed X-ray/gamma-ray emissions from PSR~B1509-58
 within framework of the outer gap model and considered that PSR~1509-58
is the Crab-like pulsar, that is, the X-ray/gamma-ray emissions are created 
by the synchrotron radiation and inverse-Compton emissions of 
the electron and positron pairs created in the outer magnetosphere. 
The calculated overall spectrum qualitatively agrees
 with the multi-wavelength data. 
However, the light cylinder radius of PSR~B1509-69 is so large as to make it 
very difficult to attenuate all of the GeV curvature photons emitted from 
the outer gap, which may be inconsistent with no detection of GeV gamma-rays 
by $Fermi$.

Harding et al. (1997) proposed operation of the photon-splitting process 
for the formation of the soft gamma-ray spectrum of PSR B1509-58, 
whose spin down dipole magnetic field is $B_s\sim 2\times 10^{13}$G. 
They argued that as the stellar magnetic field approaches the critical value, 
$B_c\sim 4.4\times 10^{13}$G, the magnetic photon-splitting process 
plays an important role as attenuation of the gamma-rays emitted 
in the polar cap region.  They discussed that the photon-splitting 
and pair-creation cascade process can explain the position of the 
spectral peak $\sim$10MeV of PSR B1509-58. 
However, this model will not explain  the soft gamma-ray spectra of 
PSRs J1617-5055 and PSR J1811-1925, whose inferred 
magnetic fields are  only $B_s\sim 3\times 10^{12}$G and $2\times 10^{12}$G, 
respectively, which are the typical values of the canonical gamma-ray 
pulsars.

 Wang et al. (2013) proposed a new model for PSR~B1509-58 in the framework 
of the outer gap accelerator model (Takata et al. 2010; Wang et al. 2010).  
They discussed that the Earth viewing angle measured from the rotation axis is 
smaller than (or close to) the inclination angle of the magnetic axis. 
In such a small viewing angle, the outward GeV emissions, which 
creates the observed spectra of the $Fermi$-LAT  pulsars,  are missed by 
the observer, while the inward emissions 
contribute to the observed emissions. Wang et al. (2013) argued furthermore  
that the magnetic pair-creation cascade initiated by 
 the inward 0.1-1GeV emissions near the stellar surface   
eventually produces the soft spectrum of the PSR B1509-58.
 Lin et al. (2009) also proposed that the X-ray emissions 
from GeV-quiet SGPSR J1838-0655 is produced by the synchrotron radiation 
of the pairs, which are produced by the magnetic pair-creation process 
of the inward gamma-rays from the outer gap. 

Main purpose of this paper is to apply the  model of the inward emissions to 
 other   GeV-quiet SGPSRs, 
since the member of the GeV-quiet SGPSRs is growing and 
 since no previous studies have been discussed the emission mechanisms. 
 In particular, we will apply our model to four GeV-quiet 
 soft gamma-ray pulsars, PSRs J1617-5055, J1811-1925, J1846-0258 
and J1930+1852, for which  detailed spectral data in 10-100keV 
bands were found in the literature.  Although 
 no detection of the emissions above 100keV  has been reported, 
 they share some properties of
 the emissions with PSR B1509-58; for example, (1) their
 radio emissions are dim or quiet, 
(2) the pulse profile in  X-ray/soft gamma-ray bands is described by
 a single broad curve, (3) there are no  GeV emissions and (4) the 
 broad band spectral shape suggests the maximum energy 
flux at MeV energy bands.  It is likely therefore that the emission processes
 of those GeV-quiet SGPSRs are 
different from the typical gamma-ray pulsars.  
 The spin down parameters of those  soft gamma-ray pulsars
 are summarized in Table~1.
 
In the paper, we also analyze the six year $Fermi$ data and determine 
the upper limit flux of the GeV emissions (section\ref{fermi}), 
because we could not find any published results.
 We describe theoretical model in section~\ref{model} and compare 
the calculated spectra and light curves in section~\ref{result}. 
A brief summary is presented in section~\ref{summary}.

\section {$Fermi$ data analysis}
\label{fermi}
We used the $\gamma$-ray data from the {\it Fermi} Large Area Telescope (LAT)
to search any gamma-ray emissions from   the four soft gamma-ray pulsars, 
PSR J1617$-$5055, PSR J1811$-$1925, PSR J1846$-$0258 and PSR J1930$+$1852. The data analysis was performed using
the {\it Fermi} Science Tools package (v9r32p5) availabe from the {\it Fermi} Science 
Support Center (FSSC) \footnote{http://fermi.gsfc.nasa.gov/ssc/data/analysis/software/}.
The data we used here were obtained from the reprocessed {\it Fermi} Pass 7 database and the
instrumental response function used was the P7REP\textunderscore SOURCE\textunderscore V15 version. 
We used the data in the period starting from 2008-08-04 15:43:37 to 2014-05-30 01:27:16 (UTC). 
We selected the photons carrying energy between 100 MeV and 100 GeV within 20$\degr$$\times$20$\degr$ 
regions of interest (ROI) centered at the positions of the pulsars. To prevent the contamination by the
Earth's albedo, the events with zenith angle greater than 100\degr\ or rocking angle greater than
52\degr\ were filtered. 
	
Binned likelihood analysis was performed using the {\it gtlike} function. To model the background
source contributions, we included all 2-year {\it Fermi} Gamma-ray LAT (2FGL) catalog point sources 
(Nolan et al., 2012) associated with the extended source templates within 20\degr from the ROI center. 
The spectral parameters for sources greater than 10\degr from the pulsars were kept fixed to the values 
defined in the catalog. For sources between 6\degr\ and 10\degr\ away from the center of ROI, only the 
spectral indices were kept fixed to the catalog definitions. The galactic diffuse background 
(gll\textunderscore iem\textunderscore v05.fits) and the isotropic diffuse background 
(iso\textunderscore source\textunderscore v05.txt) were also included in the modeling. All of these 
background modeling resources are available from the FSSC. 
	
Using the full energy range extracted, 100 MeV to 100 GeV, we modeled the four soft gamma-ray pulsars 
as point sources using the simple power law
\begin{equation}
{\textrm{d}N\over \textrm{d}E} = N_{0}\left({E\over E_{0}} \right)^{-\Gamma}. 
\end{equation}
The spectral energy distributions (SEDs) under 1 GeV were calculated using the modeled power law with 
all the spectral indices in the model kept fixed to the best-fit values. Two equally divided energy 
bins, 100 MeV to 316 MeV and 316 MeV to 1 GeV, were used in this analysis. 

\begin{figure}
\centering
\includegraphics[width=1\textwidth]{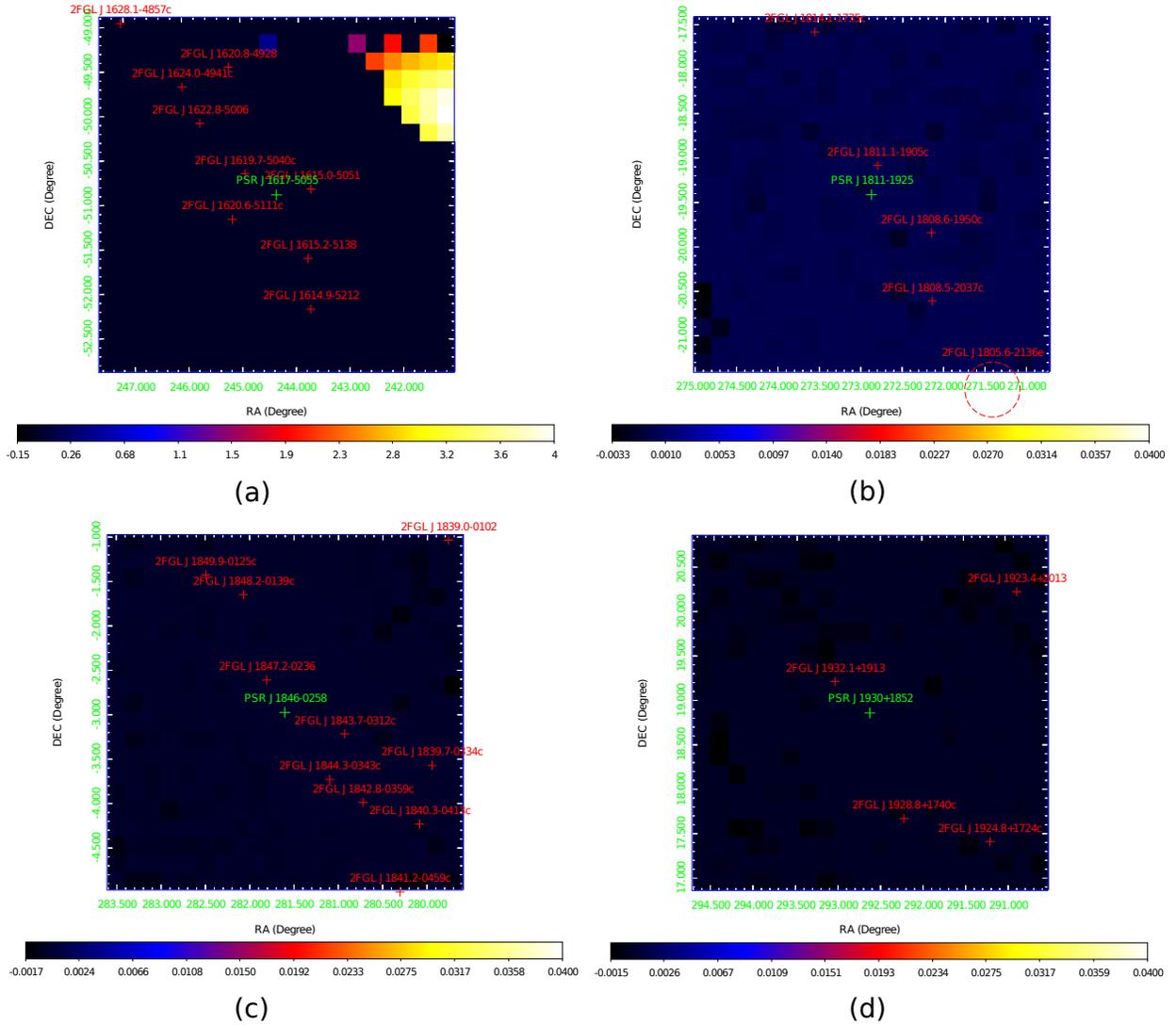}
\caption{100 MeV to 316 MeV TS maps in the 4\degr$\times$4\degr\ regions centered at
(a) PSR J1617$-$5055, (b) PSR J1811$-$1925, (c) PSR J1846$-$0258 and (d) PSR J1930$+$1852. }  
\label{fig:100to316mevTSmaps}
\end{figure}

\begin{figure}
\centering
\includegraphics[width=0.5\textwidth]{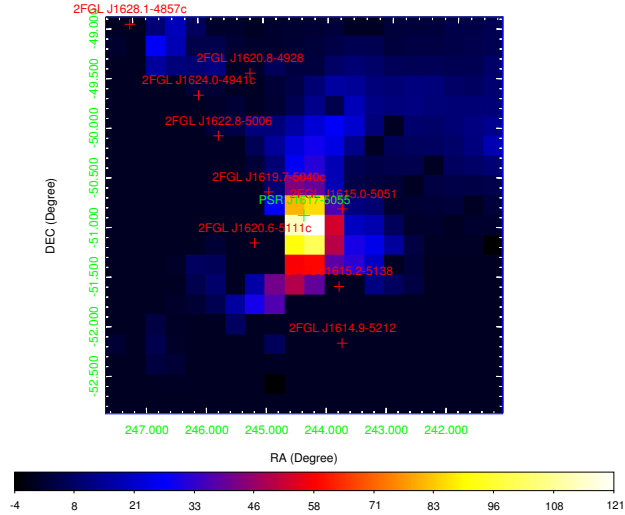}
\caption{TS-map for PSR J1617$-$5055 in energy range from 100 MeV to 100 GeV.}
\label{fig:1617TSfull}
\end{figure}

Figure ~\ref{fig:100to316mevTSmaps} shows four test-statistic (TS) maps created in the 4\degr$\times$4\degr\ regions centered at the four pulsars with energy ranged from 100 MeV to 316 MeV. The TS values indicated 
that there is no detection by the {\it Fermi}-LAT in this energy band at the locations of the soft 
gamma-ray pulsars. Only upper limits for the emissions under GeV could be determined from the LAT data. 
The values were tabulated in Table ~\ref{table:upperlimits}. It is noted that a significant source is detected
by {\it Fermi}-LAT above GeV range at the position of PSR J1617$-$5055. Figure ~\ref{fig:1617TSfull} shows a 
TS map of PSR J1617$-$5055 with energy ranged from 100 MeV to 100 GeV and TS $\simeq$ 100 (10$\sigma$) at 
the central position. This detection was reported by Xing et al. (2014) and identified as the {\it Fermi} 
$\gamma$-ray counterpart to the supernova remnant RCW 103.

\section {Theoretical  model}
\label{model}
\begin{figure}
\centering
\includegraphics[width=1.\textwidth]{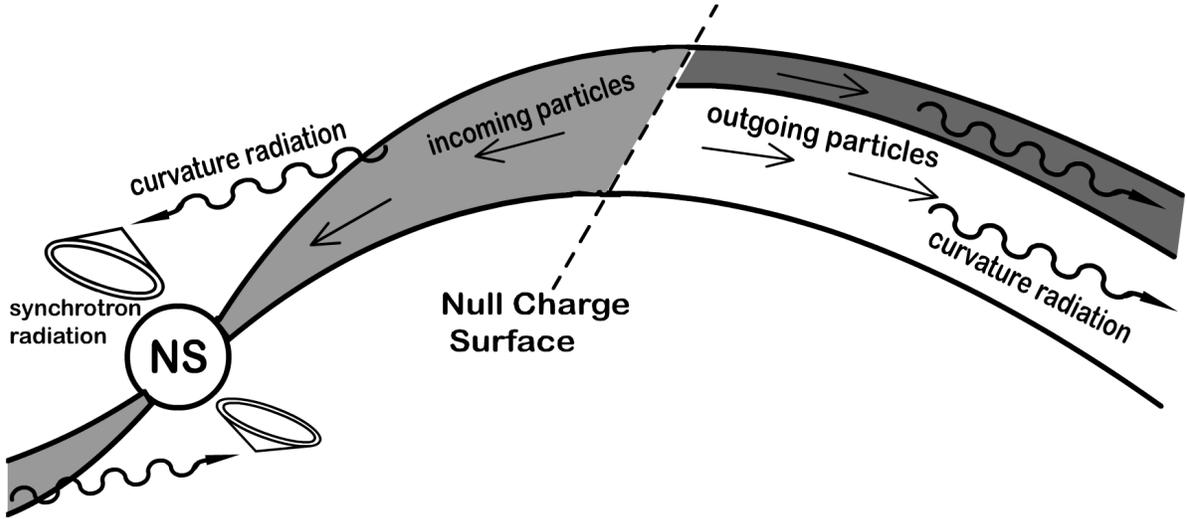}
\caption{Schematic view of the inward emissions from the outer 
gap accelerator. This figure is from Wang et al. (2013). Primary particles 
accelerated in the gap emit the gamma-ray photons via the curvature 
radiation process in the direction of its motion, which comprises of 
the motion along the magnetic field line and the co-rotation motion.  
 Incoming particles  emit the  100MeV-1GeV gamma-rays below the 
null charge surface. The 100MeV-1GeV gamma-rays emitted near 
the stellar surface interacts with the strong magnetic field and 
produce new pairs. The gyration motion of the pairs with an infinite pitch 
angle  produces hard X-ray/soft gamma-rays 
via the synchrotron radiation, which covers
 a wider sky area  than the curvature radiation of the primary particles}
\label{struc}
\end{figure}

\begin{table}
\begin{tabular}{cccccccc}
\hline
PSRs & $P$ & $L_{sd}$ & $B_{s}$ & $D$ & $\alpha$ & $\beta$ & $f_{gap}$ \\
 & (ms) & ($10^{36}{\rm erg s^{-1}}$) & ($10^{12}$G)& (kpc)
 & (degree) & (degree) &  \\
\hline\hline
B1509-58 & 151 & 17 & 15 &4.4  & 30 & 15 & 0.3 \\
J1617-5055 & 69 & 16 & 3.1 & 6.5 &15 & 25 & 0.21 \\
J1811-1925 & 65 & 6.4 & 1.7 & 5&10 & 35 & 0.2 \\
J1838-0655 & 65 & 6.4 & 1.7 & 6.6 &- & - & - \\
J1846-0258 & 326 & 8.1 & 48 & 5.8& 10 & 35 &  0.5 \\
J1930+1852 & 136 & 12 & 10 & 5 &20  & 41 & 0.28 \\
\hline
\end{tabular}
\caption{GeV-quiet soft gamma-ray pulsars. 
The second ($P$), third $(L_{sd})$ and 
fourth ($B_{s}$) columns are rotation period, spin down power and 
surface dipole magnetic field, respectively. The fifth ($D$) is 
the distance to the source and is used to estimate 
the observed luminosity in Figure~\ref{sgr}.
 The sixth ($\alpha$), seven ($\beta$) and eight ($f_{gap}$) are the magnetic 
inclination angle, Earth viewing angle and the fractional gap thickness, 
respectively, inferred from the fitting of the observations. 
The fitting result for PSR B1509-68 was taken from Wang et al. (2013). 
We did not fit PSR~J1838-0655, since no  point of the spectral data 
for the hard X-ray emissions ($>$10keV) from the pulsar 
have not been published.}  
\end{table}
\begin{table}
\begin{tabular}{lrr}
\hline
PSRs & Emission in 100 MeV -- 316 MeV & Emission in 316 MeV -- 1 GeV\\
& (erg cm$^{-2}$ s$^{-1}$) & (erg cm$^{-2}$ s$^{-1}$) \\
\hline
\hline
J1617$-$5055 & $<$2.5 $\times 10^{-11}$ &  $<$9.1 $\times 10^{-12}$ \\
J1811$-$1925 & $<$3.7 $\times 10^{-12}$ &  $<$4.9 $\times 10^{-12}$ \\
J1846$-$0258 & $<$4.4 $\times 10^{-11}$ &  $<$1.0 $\times 10^{-11}$ \\
J1930$+$1852 & $<$1.9 $\times 10^{-11}$ &  $<$5.7 $\times 10^{-12}$ \\
\hline
\end{tabular}
\caption{Upper limits for the pulsars PSR J1617$-$5055, PSR J1811$-$1925, PSR J1846$-$0258 and 
PSR J1930$+$1852 in the energy bands: 100 MeV -- 316 MeV and 316 MeV -- 1 GeV.
\label{table:upperlimits}}
\end{table} 
In our model, we suggest the emissions from the GeV-quiet 
soft gamma-ray pulsars are produced via synchrotron radiation 
of the pairs, which are created by the interaction of 
the inward gamma-rays and the strong magnetic field near the polar cap region. 
 Figure~\ref{struc} shows schematic picture for the inward gamma-rays 
emissions and subsequent pair-creation process and synchrotron 
radiation process.  We emphasize that the viewing geometry 
is a crucial factor to differentiate between the typical gamma-ray pulsars 
and soft gamma-ray pulsars. Our model  expects that  
 {\it millisecond} pulsar does not show GeV-quiet soft gamma-ray spectrum,
 since  its dipole magnetic field $B_s\sim 10^{8-9}$G is too small 
to operate the magnetic pair-creation process 
(except for very close to stellar surface, where 
the stronger multi-pole  magnetic field may dominate the dipole field). 
 Since a detail method of the calculation was described by Wang et al. (2013), 
we briefly mention the guideline of the model. 

\subsection{Inward emissions of the outer gap}
In the outer gap model, the charged particles are accelerated 
by the electric field parallel to the magnetic field, and 
emit the GeV gamma-rays via the curvature radiation process. 
Takata et al. (2008) argued that the outer gap accelerator 
produces {\it outward} and {\it inward} gamma-rays, 
which are produced by the outgoing particles and incoming particles 
accelerated in the gap; for the magnetic  inclination angle smaller 
than 90~degree, the outgoing and incoming particles are 
positrons and electrons, respectively.  In the outer gap, 
since (i) the strong acceleration region extends 
between the null charge surface of the Goldreich-Julian charge density 
and the light cylinder and
 (ii) most of pairs are produced {\it around  the null charge
 surface} (Cheng et al. 2000), 
the outgoing particles are accelerated by almost full potential 
drop in the gap, while the incoming particles 
feel only potential drop  between the inner boundary and 
the pair-creation position. Hence, it is expected that
 the luminosity of the outward propagating 
gamma-rays are about one order of magnitude larger than that of inward 
propagating gamma-rays, suggesting the $Fermi$ has preferentially detected 
the outward emissions of the outer gap. Within the framework of 
 the outer gap model, the gamma-ray luminosity can be written as 
\begin{equation}
L_{\gamma}\sim I_{gap}V_{gap},
\end{equation}
where $I_{gap}$ is total current in the outer gap and $V_{gap}$ is electric
 potential drop along the magnetic field line. As aforementioned pair-creation 
region in the outer gap accelerator, 
the outgoing particles are accelerated by almost full potential 
drop in the gap,  which can be estimated as 
\begin{equation}
V_{gap}^{out}\sim  f^2_{gap}B_{lc}R_{lc}, 
\end{equation}
where $R_{lc}=Pc/2\pi$ is the light cylinder radius, $B_{lc}$ is the magnetic 
field at the light cylinder, and  $f_{gap}$, which takes a value
 of $\sim 0.2-0.3$, is the ratio of the gap thickness and the light
 cylinder radius at the light cylinder. For the inward emissions, 
the incoming particles are accelerated with a potential of  
$V_{gap}^{in}\sim 0.1V_{gap}^{out}$. 

In the outer gap magnetosphere, the charge particles are accelerated 
by the electric field along the magnetic  field line 
$E_{||}\sim V_{gap}/R_{lc}$
 and emit gamma-rays through the curvature radiation process.  
Assuming balance between the electric force and radiation drag force, 
the saturated Lorentz factor is proportional to $\Gamma\propto V_{gap}^{1/4}$. 
As a result, the typical energy of the curvature radiation is 
proportional to $E_c\propto \Gamma^{3}\propto V_{gap}^{3/4}$. Since 
$V_{gap}^{in}\sim 0.1V_{gap}^{out}$, the  energy of the curvature radiation 
of inward emissions is a factor of $\sim 5$ smaller than that 
of the outward emissions and it typically becomes  0.1-1GeV.

For the outer gap accelerator, the strong acceleration region extends beyond the null charge surface, which is defined
 by surface of 
$\mathbf{\Omega}\cdot\mathbf{B}=0$.  
It has been proposed that the active 
outer gap with the electric current can be extended ``below''
 null charge surface 
(Takata et al. 2004; Hirotani 2006)), but the accelerating electric 
field below the null charge surface is significantly reduced by the electron
 and positron pairs with a very weak field. We approximate the electric 
structure below the null charge surface as 
\begin{equation}
E_{||}(r<r_{null})=\frac{(r/r_{in})^2-1}{(r_{null}/r_{in})^2-1}E_{||,null},
\end{equation}
where $E_{||,null}$ is the electric field strength at the null charge 
surface and is given by our three-dimensional two-layer structure 
model (Wang et al. 2011), and $r_{null}$ is the radial distance to the null charge surface, which is a function of the inclination angle and azimuth angle. 
In addition, $r_{in}$ is 
the radial distance to the inner boundary of the outer gap and is set at 
$20$ stellar radius. 

Near and below the inner boundary, the incoming particles loose their 
energy via the curvature radiation process.  
When the Lorentz factor of the incoming particles drops low enough,
 the curvature energy loss time scale becomes comparable to the time scale
 of the particle's movement to the stellar surface. In such a case, 
we can show that the energy of the curvature photon is 
 $9m_ec^2/8\alpha_f\sim$100 MeV, where $\alpha$ is the fine structure constant
 (Takata et al. 2010). Hence, we expect 
that the incoming particles emit 0.1-1GeV photons between the null charge 
surface and the stellar surface.  The gamma-ray  photons emitted below 
the null charge surface may pass 
through the strong magnetic field region near the stellar surface 
and may initiate magnetic pair-creation cascade.

\subsection{Magnetic pair-creation cascade}
The typical cut-off energy  0.1-1GeV in the spectrum of the inward emissions 
will be still higher than the spectral cut-off energy ($\sim$1-10MeV) 
 of the GeV-quiet SGPSRs; for example, 
the original soft gamma-ray pulsar, PSR B1509-58, shows a spectral cut-off 
at $\sim 5$MeV. To explain the position of the spectral cut-off 
of PSR B1509-58,  we simulate the pair-creation cascades of 
the inward gamma-ray emissions (Wang et al. 2013). 
If the inward propagating gamma-rays emitted below the null charge surface  
 pass through near the stellar surface,
 they may be absorbed by 
the magnetic field and be converted into electron and 
positron pairs (magnetic pair-creation process). The mean free path
 of the magnetic pair-creation may be written as (Erber 1966)
\begin{equation}
\ell =\frac{4.4}{(e^2/\hbar c)}\frac{\hbar}{m_ec}\frac{B_c}{B_\perp}
\mathrm{exp}\left(\frac{4}{3\chi}\right),
\end{equation}
where $\chi=\hbar\omega B_{\perp}/(2m_ec^2B_c)$ 
and $B_{\perp}=B\sin\theta_{p}$ with $\theta_{p}$ being the angle between 
the magnetic field direction and propagating direction of the photon
 and $B_c=4.4\times 10^{13}$G.  We calculate the optical depth 
$\tau_{opt}(s_i)=\int_{s_{i-1}}^{s_{i}} ds/\ell(s)~(i=1,2,3..)$, where $s_0=0$ 
corresponds to  the position of the emitted point. We 
 determine the pair-creation position $s_i$ from the condition
 $\tau(s_{i+1})-\tau(s_i)=0.1$ and calculate the number of created 
pairs from  the equation  $\delta N(s_i)=N_0\{{\rm exp}[-\tau(s_{i-1})]-
{\rm exp}[-\tau(s_i)]\}$,  where $N_0$ 
is the emitted gamma-rays in the gap.
 We also taken into account the pair-creation 
process of the gamma-rays with the X-rays.

\subsection{Synchrotron emissions from new pairs}
\label{synch}
The created pairs have a pitch angle $\theta_p$ and loose their-energy 
via the synchrotron radiation. We solve the evolution of the Lorentz factor 
$(\gamma$) of the pairs with the equations of 
\begin{equation}
\frac{dP_{||}}{dt}=-\frac{2e^4B^2\gamma^2\sin^2\theta_p}{3m^2c^4}\cos\theta_p
\end{equation}
and 
\begin{equation}
\frac{dP_{\perp}}{dt}=-\frac{2e^4B^2\gamma^2\sin^2\theta_p}{3m^2c^4}\sin\theta_p,
\end{equation}
where $P_{||}=m_ec\gamma\cos\theta_p$ and $P_{\perp}=m_ec\gamma\sin\theta_p$. 
Since the magnetic field and Lorentz factor of the particle at the 
pair-creation position are $B_{\perp}=2m_ec^2B_c/(\chi E_{\gamma})$ and 
$\gamma=E_{\gamma}/2m_ec^2$, respectively,  
the maximum energy of the synchrotron 
radiation of the new born pairs becomes as 
\begin{equation}
E_{syn,max}\sim \frac{3\hbar \gamma^2eB_{\perp}}{2m_ec}
\sim \frac{3E_{\gamma}}{4\chi}\sim 38\left(\frac{E_{\gamma}}{0.5{\rm GeV}}\right)
\left(\frac{\chi}{0.1}\right)^{-1}{\rm MeV},  
\end{equation}
suggesting the spectrum of the synchrotron radiations of the pairs, which 
are produced by the magnetic pair-creation process, has a spectral 
turn over around 10MeV. The position of this spectral turnover can explain 
that of the GeV-quiet SGPSRs. Therefore, we suggest that the observed
 high-energy emissions from GeV-quiet SGPSRs are produced through  
the synchrotron radiation occurred near the stellar surface. We also take 
into account the magnetic pair-creation process of the synchrotron photons,
 which was ignored in Wang et al. (2013).

We take into account the effects of the pitch angle and the gyration motion 
on the emission direction of synchrotron radiation. The 
particle motion is expressed by sum of 
the motion along the magnetic field line, gyration motion and 
co-rotation motion. Taking $z$-axis along the rotation axis, the 
particle motion is calculated from (Takata et al. 2007; Wang et al. 2013)  
\begin{equation}
\mathbf{v}=\lambda v_p\mathbf{v}_{syn}'/|\mathbf{v}'_{syn}|
+\mathbf{\Omega}\times\mathbf{r},
\end{equation}
where $v_p$ is calculated from the condition $|\mathbf{v}|=c$ and 
$\mathbf{v}_{syn}'$ is given by 
\begin{eqnarray}
v'_{syn,x}&=&\hat{B}_x+\tan\theta_p(u_x\cos T+v_x\sin T)\nonumber \\
v'_{syn,y}&=&\hat{B}_y+\tan\theta_p(u_y\cos T+v_y\sin T),\\
v'_{syn,z}&=&\hat{B}_z+\tan\theta_p(u_z\cos T+v_z\sin T)\nonumber 
\end{eqnarray}
where $\theta_p$ is the pitch angle of the created pairs and $T$ is 
the phase of the gyration motion.   
In the equation above, $\hat \mathbf{B}=\mathbf{B}/|B|$, 
$\mathbf{u}=[B_y/(B_x^2+B_y^2)]^{1/2},-B_x/(B_x^2+B_y^2)^{1/2},0)$ and 
$\mathbf{v}=(\mathbf{B}\times \mathbf{u})/|\mathbf{B}\times \mathbf{u}|$.
 In addition, $\lambda$ represents the direction of the particle 
motion projected to the magnetic field line, it takes   
$\lambda=1$ for $\theta_p\le 90^{\circ}$ an $\lambda=-1$ 
for $\theta_p>90^{\circ}$. 

The Earth viewing angle $(\beta)$ measured from the rotation axis 
and the pulse phase $\psi$ for a synchrotron (or curvature) photon 
can be calculated from
\begin{equation}
\cos\beta=\frac{v_z}{\mathbf{v}}
\end{equation}
and
\begin{equation}
\psi=-\cos^{-1}(v_x/\sqrt{v_x^2+v_y^2})-\frac{\mathbf{r}\cdot \mathbf{v}}
{vR_{lc}},
\end{equation}
respectively, where $\mathbf{r}$ is the vector to the radiation point.
 For each viewing angle $\beta$, we calculate the phase-averaged spectrum 
and compare the result with the observations (section~\ref{result}).

 The synchrotron 
emissions from the pairs $(\theta_p\ne0)$ 
 with the gyration motion covers
 a wider sky area  than the curvature radiation of 
the primary particles $(\theta_p=0)$,
 which is emitted along the magnetic field line.  
Wang et al. (2013) discussed  the evolution of the X-ray/gamma-ray spectrum 
for the different viewing geometry: Fixing magnetic inclination angle at 
$\alpha=20$degree, the {\it outward curvature emissions}
  dominate the inward emissions  
and the spectrum extends up to several GeV if the viewing angle 
is  $\beta\sim 70-90$~degree.  For  mildly viewing angle $\beta\sim 50$~degree, 
the inward curvature emissions and subsequent synchrotron radiation 
of the pairs can contribute to the spectrum. 
For small inclination angle $\beta\sim \alpha$ or $\beta< \alpha$, 
only synchrotron radiation of the pairs created by the magnetic pair-creation
 contributes to the observations, 
 and the spectral peak in $\nu F_{\nu}$ appears at
 around $\sim$1MeV, which can explain the spectral properties of 
GeV-quiet SGPSRs.  Hence, our model suggests that the  GeV-quiet
 SGPSR is a peculiar case of the viewing geometry and it has a  
relatively small viewing angle and inclination angle comparing with
 the normal gamma-ray pulsars.

In the present calculation, we apply the rotating dipole magnetic field 
in vacuum (Cheng et al. 2000). Force-free magnetosphere has been investigated 
for magnetic field and current structure in the pulsar magnetosphere
 (Contopoulos et al. 1999; Spitkovsky 2006),
 and provides a distinct GeV pulse profile from the 
vacuum dipole field  (Bai \& Spitkovsky 2010). More realistic pulsar 
magnetosphere will be between the vacuum 
dipole field and the fore-free
 field (e.g. Li et al. 2012; Kalapotharakos et al. 2012). In the present 
calculation, however, since the emission regions 
in X-ray/soft gamma-ray bands are near the neutron star surface, where 
 the force-free field and vacuum dipole field may be  close to each other,
 the rotating vacuum dipole field may provide a good approximation to discuss 
the pulse profile. 
 
\section{Results}
To fit the observed spectrum, main model parameters are the fractional 
gap thickness $f_{gap}$, the magnetic inclination angle $\alpha$ and 
the Earth viewing angle $\beta$. The inclination angle and the 
viewing angle of a pulsar is sometimes constrained by rotating vector model, 
which fits the observations 
of the radio polarization (Radhakrishnan \& Cooke 1969).
For instance,  the fitting of the original soft gamma-ray pulsar, PSR B1509-58, 
suggests the inclination angle of $\alpha<60$~degree (Crawford et al. 2001).
 The geometrical model of the pulsar wind nebula  is also used to constrain the 
viewing geometry of the pulsar (Gaensler et al, 2002; Ng \& Romani 2008).
Using CHANDRA data, for example, Lu et al. (2002) found  clear torus 
structure of the pulsar wind nebula surrounding the soft gamma-ray pulsar 
PSR~B1930+1852 and suggested  the Earth viewing angle of $\beta=41$~degree.
We emphasize that within the framework of our model, the viewing geometry 
that  explains (1) the softness of the spectra, (2) it's flux level 
 and (3) the single peak in  the light curve of the GeV-quiet soft 
gamma-ray pulsars are 
constrained in a narrow range of the parameters.   In Table~1, 
we tabulate the best fitting  parameters of the 
inclination angle and the Earth viewing angle for four GeV-quiet SGPSRs.  

\label{result}
\subsection{J1617-5055}
\begin{figure}
\begin{center}
\includegraphics[height=7cm,width=7cm]{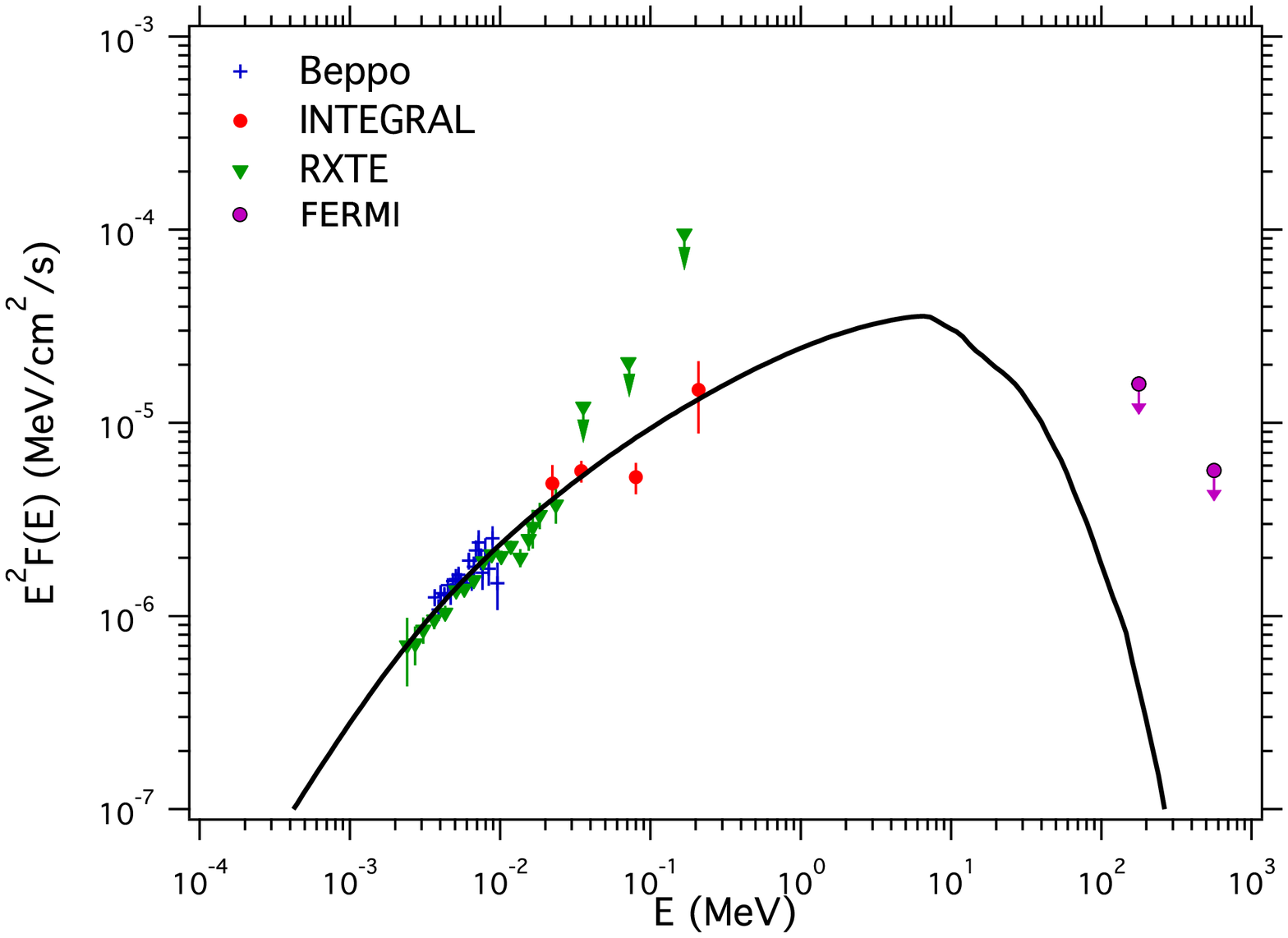}
\includegraphics[height=7cm,width=7cm]{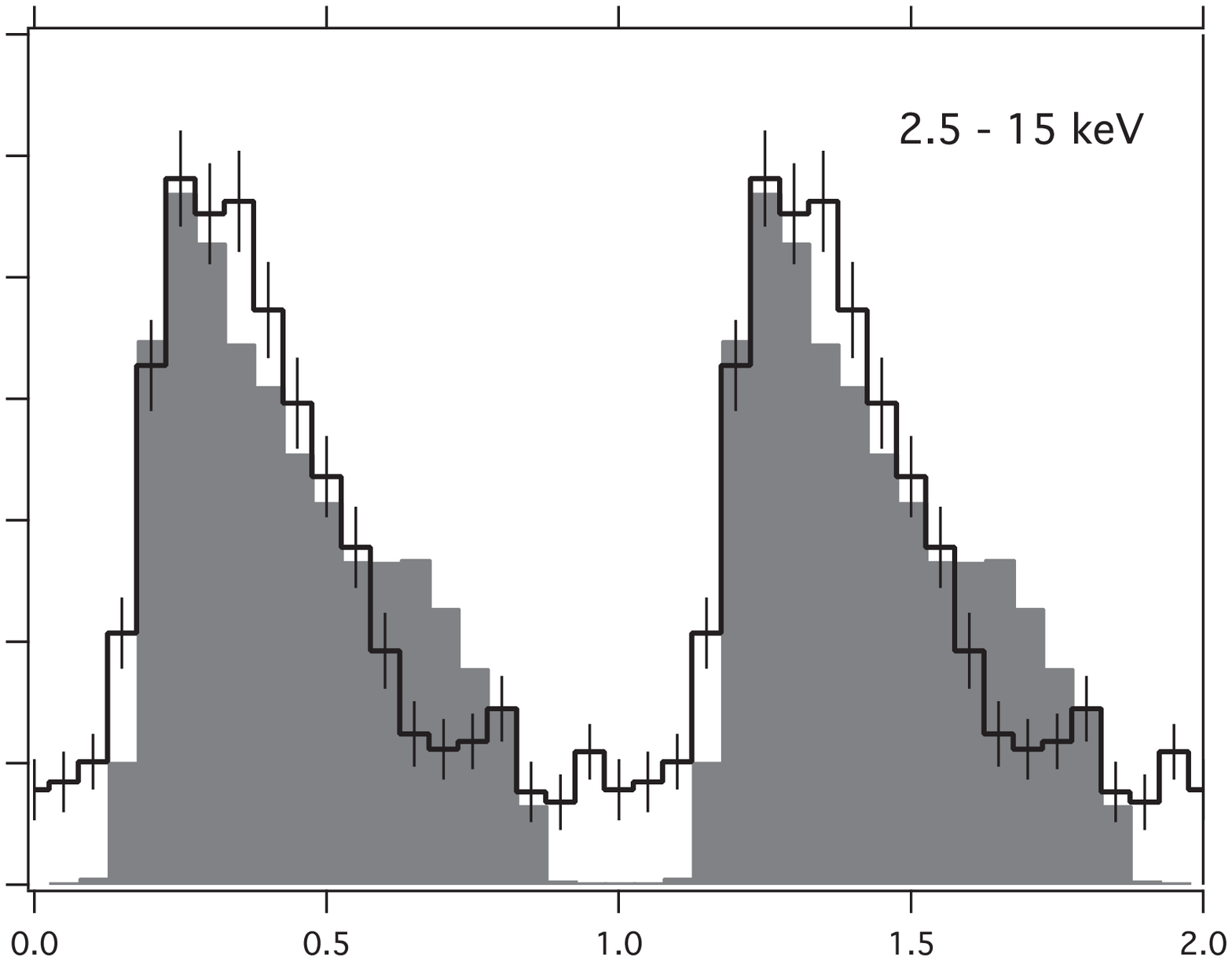}
\caption{The spectrum (left) and X-ray light curve of PSR J1617-5055.
 The solid line in the left panel and the grey histogram in the right 
panel show the calculated spectrum and the light curve, respectively. 
The results are for the inclination angle of  $\alpha=15$ degree and 
the Earth viewing angle of $\beta=25$degree. For the observed flux, 
the data were taken from Landie et al. (2007) for BeppoSAX and INTEGRAL 
and from Kuiper and Hermsen (2014) for RXTE. The upper limit of $Fermi$ 
data were determined by the this study. For the light curve, 
the data was taken from Becker and Aschenback (2002).}
\label{1617}
\end{center}
\end{figure}
The 69 ms spin-down powered pulsar PSR J1617-5055 was discovered 
by the  X-ray observations 
(Torii et al. 1998, Garmire et al. 1999; Becker \& Aschenbach 20002; Kargaltsev et al. 2009)
  with the  radio pulsation founded  shortly 
afterwards (Kaspi et al. 1998). Soft gamma-ray emissions at $\sim 100$keV
 bands were discovered by INTEGRAL (Landi et al. 2007). 
 The timing analyses show that the 
spin down dipole magnetic field is $B_s\sim 3\times 10^{13}$Gauss and the 
characteristic age is $\tau_a\sim 8.1$kyr. 
The dispersion measure gives the distance to 
the pulsar of $d\sim 6.1-6.9$kpc. The X-ray spectrum below $\sim 10$keV 
is fitted by a single power law function with a photon index of 
 $p\sim 1.4$, and there is a spectral break around 10keV 
(Torii et al. 1998; Becker and Aschenback 2002). The inferred X-ray 
conversion efficiency in 0.5-10keV bands is 
$L_X/\dot {E}\sim 1.4\times 10^{-3}$ for a distance of $d=6$kpc 
(Becker and Aschenback 2002).  These X-ray properties are common among 
the GeV-quiet soft gamma-ray pulsar. Both the X-ray and radio pule 
profiles of PSR J1617-5055 show a single peak, but absolute phase 
difference between the X-ray and radio peaks has not been known. 

Figure~\ref{1617} compares the calculated spectrum (left panel, solid line) and 
X-ray light curve (right panel, grey histogram) with the observations; 
the phase 0 (and 0.5) in Figure~\ref{1617} corresponds to the phase at which 
 the magnetic axis points towards the Earth. 
We assumed the inclination angle of $\alpha=15$~degree and the Earth viewing 
angle measured from the rotation axis of $\beta=25$~degree.  
In the present 
scenario, we have argued that the emissions from the GeV-quiet SGPSRs  
are created  by the synchrotron radiation process of the pairs, which 
are produced  by the interaction of the inwardly emitted GeV gamma-rays 
and the strong magnetic field near the pulsar. Since the pairs are mainly 
produced above the polar cap, the pulse peak with a strong 
 the synchrotron emission appears if the line of sight cuts near 
the polar cap region.
 Hence it is  required the condition that the earth viewing angle is not 
significantly shifted from the magnetic inclination angle. Furthermore, 
a small inclination angle is required to avoid the detection of 
curvature radiation (GeV emissions) from the outer gap.

\subsection{J1811-1925}
\begin{figure}
\begin{center}
\includegraphics[height=7cm,width=7cm]{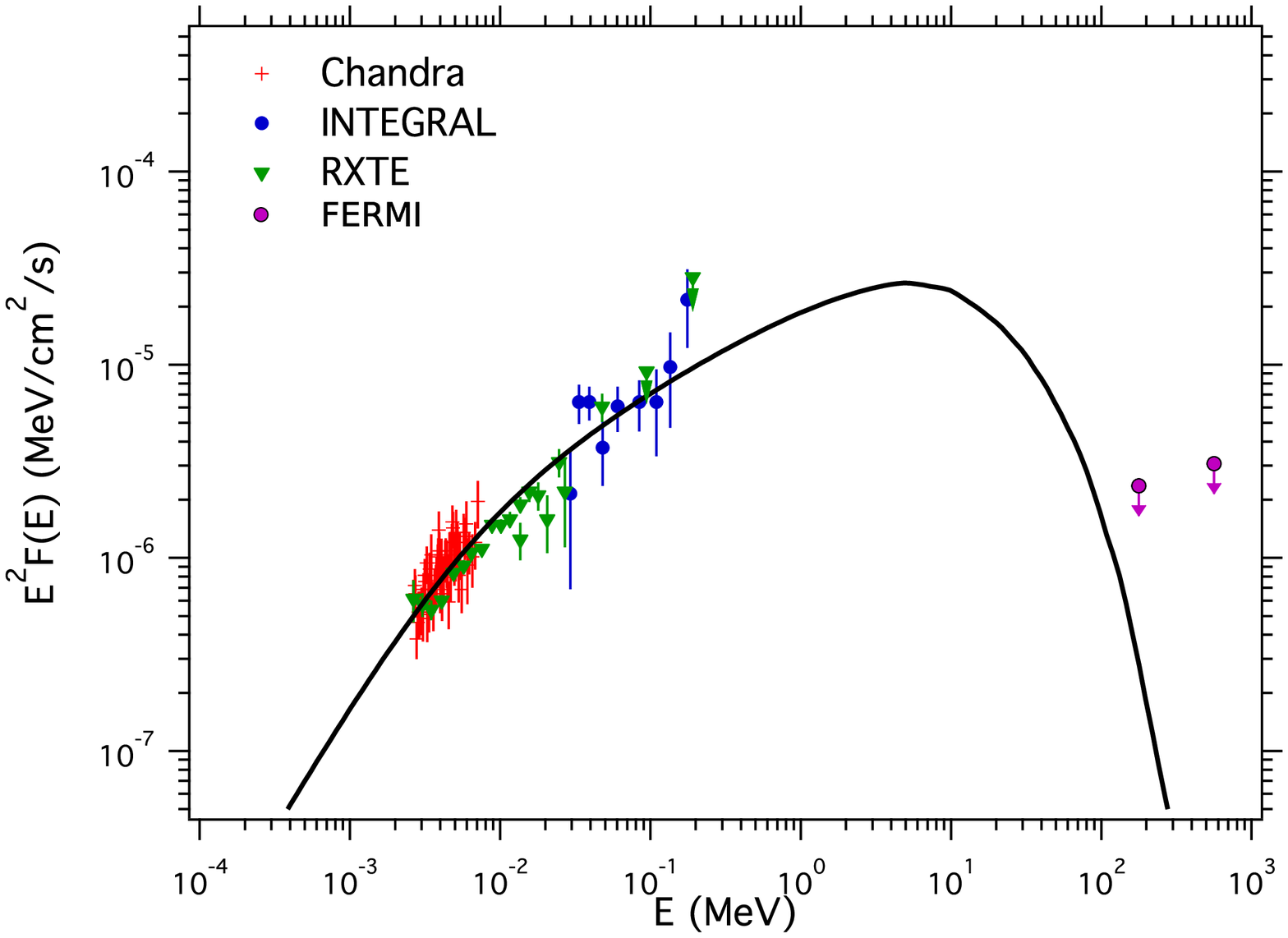}
\includegraphics[height=7cm,width=7cm]{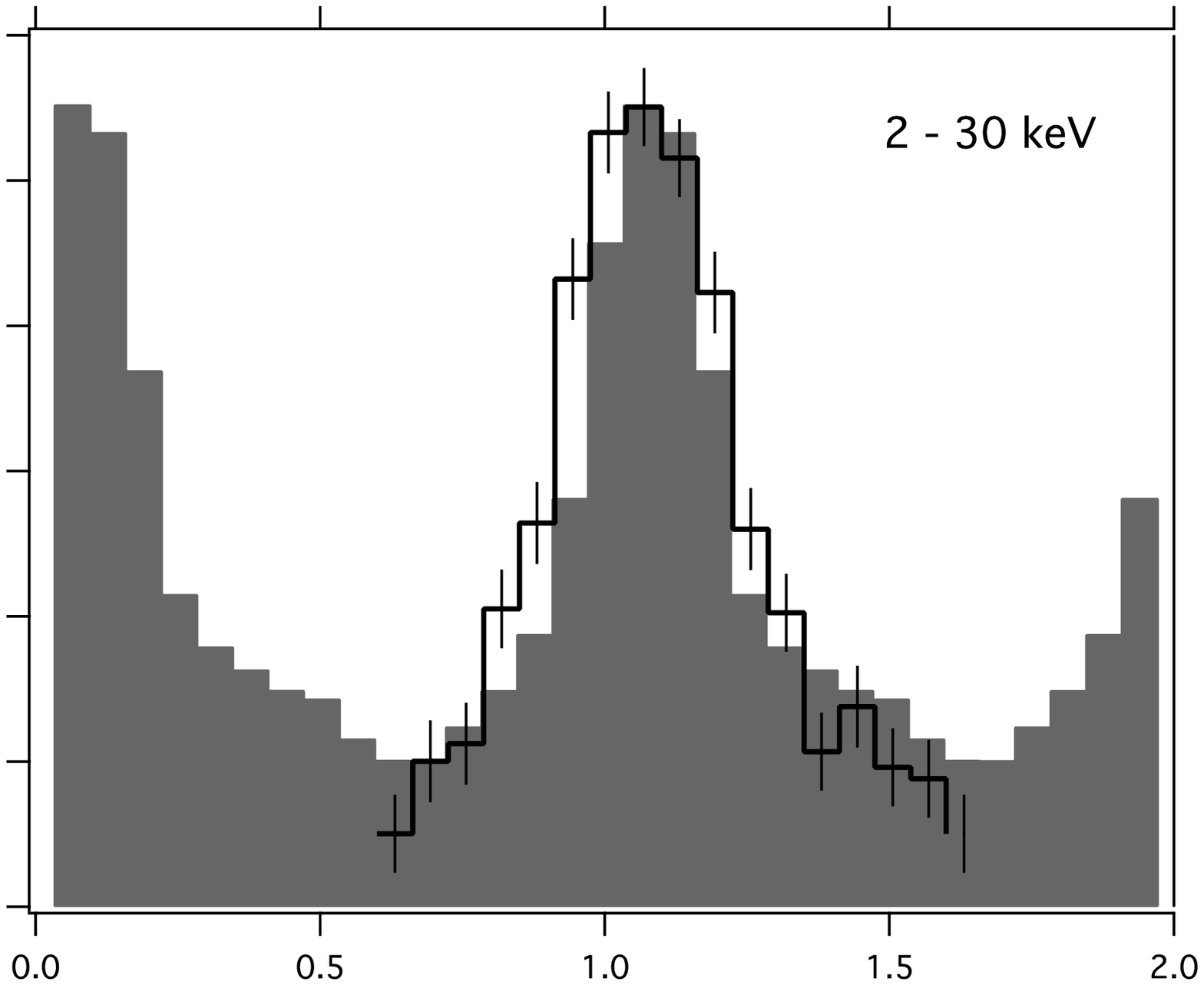}
\caption{Same as with Figure~\ref{1617}, but for PSR J1811-1925. 
The model calculation is result for $\alpha=10$degree and $\beta=35$degree, 
respectively. For the observed flux, the data were taken from 
Dean et al. (2008) for Chandra and INTEGRAL and from Kuiper and Hermsen (2014) 
for RXTE. For the light curve, the data were taken from Gavriil et al. (2004).
 Although the pulsed component above 10keV was not discussed in 
 Dean et al. (2008), it may dominate the nebula component above 10keV 
(c.f. Figure~2 in Dean et al. 2008). }
\label{1811}
\end{center}
\end{figure}

The 65-ms pulsar PSR J1811-1925 at the center of G11.2-0.3 
was discovered by  ASCA observations (Torii et al. 1997; Kaspi et al. 2001).
 This pulsar has not been detected in the radio band (Crawford et al 1998). 
The X-ray timing analysis suggests that the spin down dipole 
magnetic field is $B_s\sim 2\times 10^{12}$G and 
spin-down age is $\tau\sim 24$kyr (Torii et al. 1999). However, the CHANDRA
 observation combined with Very Large Array observations (Roberts et al. 2003)
 suggests that the reverse shock of SNR has not yet reached the PWN, 
which indicates that the system is about 2000 years old, which is consistent 
with the historical record of supernova in A.D. 386 (Clark \& Stephenson 1977).
 The distance to the pulsar is $d\sim 5$kpc as inferred from HI measurements
 (Becker et al. 1985; Green et al. 1988). Figure~\ref{1811} compares
the calculated spectrum (left panel, solid line) and 
X-ray light curve (right panel, grey histogram) with the observations. 
We assumed the inclination angle of $\alpha=10$degree and the Earth viewing 
angle measured from the rotation axis of $\beta=35$degree.

\subsection{J1846-0258}

\begin{figure}
\begin{center}
\includegraphics[height=7cm,width=7cm]{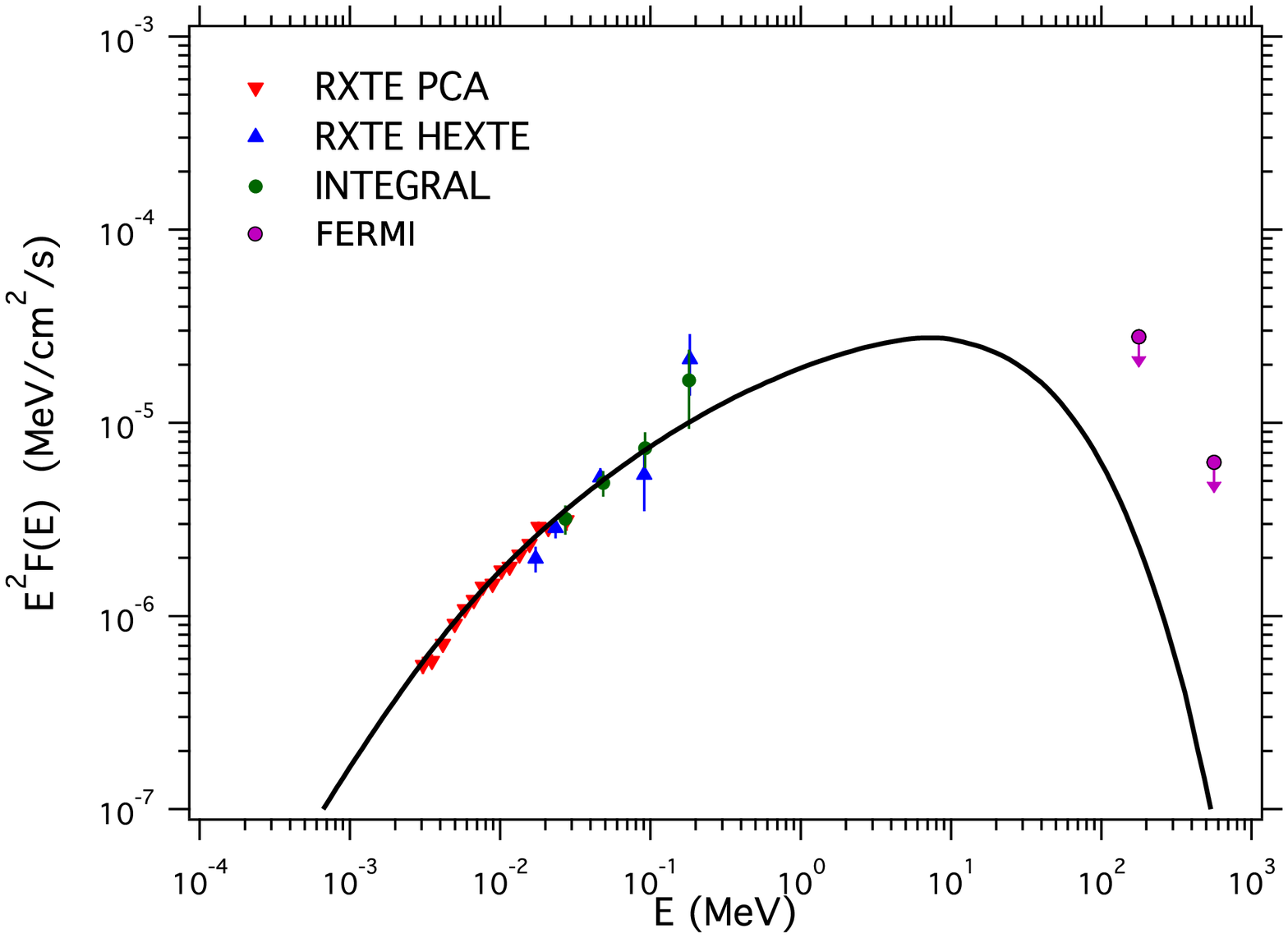}
\includegraphics[height=7cm,width=7cm]{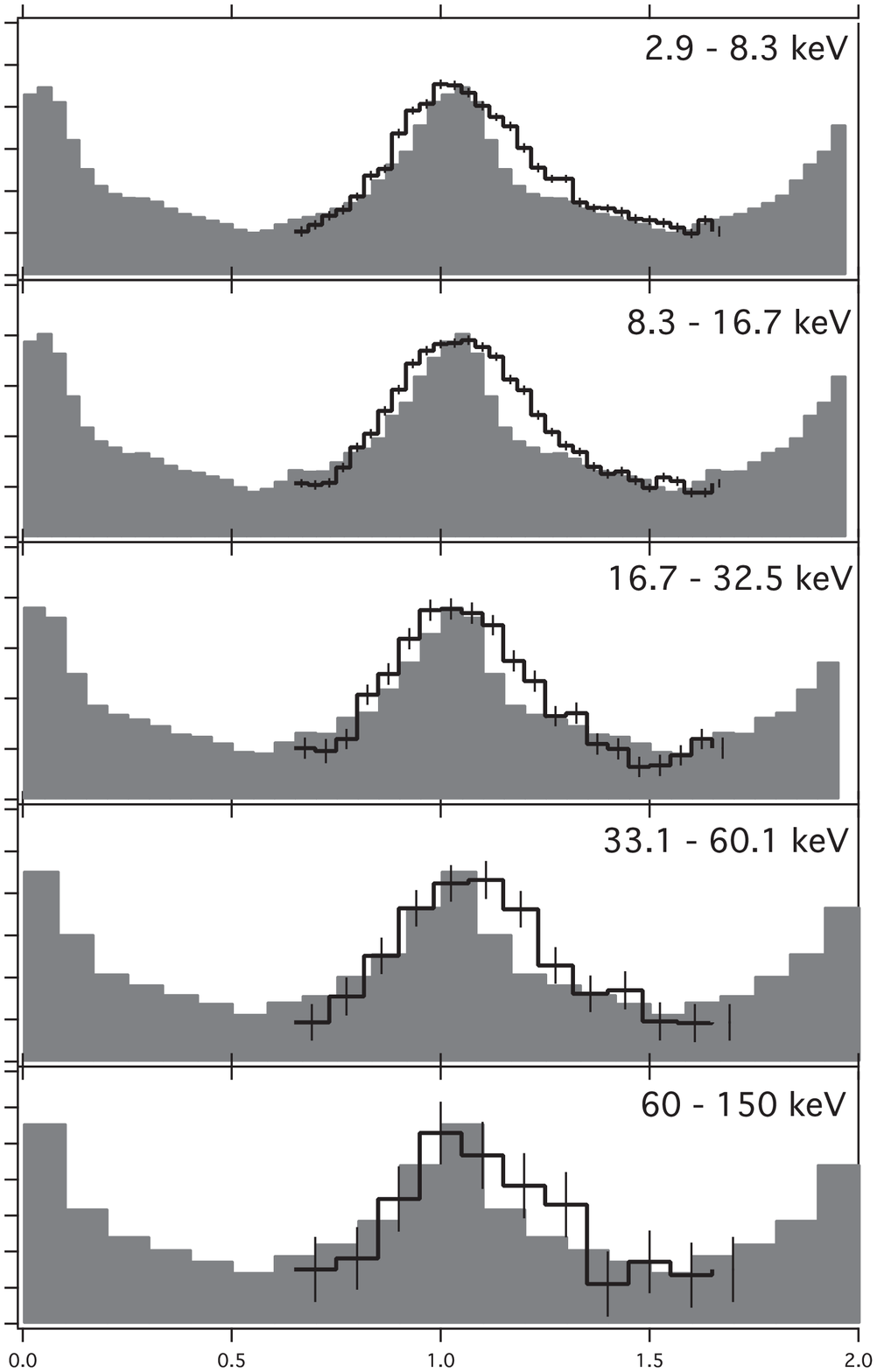}
\caption{Same as with Figure~\ref{1617}, but for PSR J1846-0258. 
The model calculation is result for $\alpha=10$degree and $\beta=35$degree, 
respectively. For the observed flux and light curve in X-ray/soft 
gamma-ray bands, the data were taken from Kuiper and Hermsen (2014).}
\label{1846}
\end{center}
\end{figure}
The 326 ms pulsar PSR J1846-0258 (also known as AX J1846.4-0258)
 was discovered by Gotthelf et al. (2000) in the X-ray bands, and is at 
the center of SNR Kes~75 (c.f. Kesteven 1968; Helfand et al. 2003; 
Molkov et al. 2004; Bird et al. 2007; Kumar \& Safi-Harb 2008; 
McBride et al. 2008;  Ng et al. 2008). No radio emission has been observed 
from PSR~J1846-0258 (Archibald et al. 2008), 
but the X-ray timing analysis reveals the surface dipole magnetic field 
of $B_s\sim 4.8\times 10^{13}$Gauss and the characteristics 
age of $\tau\sim 728$yr. It is thought that PSR~18460-258 is a transition 
object between rotation-powered pulsar and magnetically  powered pulsar 
(i.e. {\it magnetar}). In fact, the pulsar showed 
  a magnetar-like X-ray outburst accompanied by a large glitch 
in 2006 (Gavriil eta l. 2008). 

Parent et al. (2011) derived the upper limit 
of the pulsed gamma-ray flux at  
$3\times 10^{-11}{\rm erg cm^{-2} s^{-1}}$, which will be  consistent with 
the result of our $Fermi$ data analysis (c.f. Table~2 and Figure~\ref{1846}).
The distance to the pulsar is the controversial in the range of $d=5-19$kpc 
(Becker and Helfand 1984; Leahy and Tian 2008; Su et al. 2009). 
It is also known that the pulsar 
is embedded in a pulsar wind nebula (Helfand et al. 2003; McBride et al. 2008). 
Modeling of the torus in the PWN suggests a line of sight angle 
$\beta\sim 60$~degree (Ng \& Romani 2008). In our model, on the other hand, 
we require a smaller viewing angle $\beta\sim 35$~degree
 with $\alpha\sim 10$~degree to reproduce the observed emission properties.  

Figure~\ref{1846} compares  
the calculated spectrum (left panel, solid line) and 
X-ray light curve (right panel, grey histogram) with the observations. 
 In the figure, we show the light curves in the different energy bands. 
The observations suggest a single broad peak in X-ray/soft gamma-ray bands, which can be explained by 
the present model.

\subsection{J1930+1852}
\begin{figure}
\begin{center}
\includegraphics[height=7cm,width=7cm]{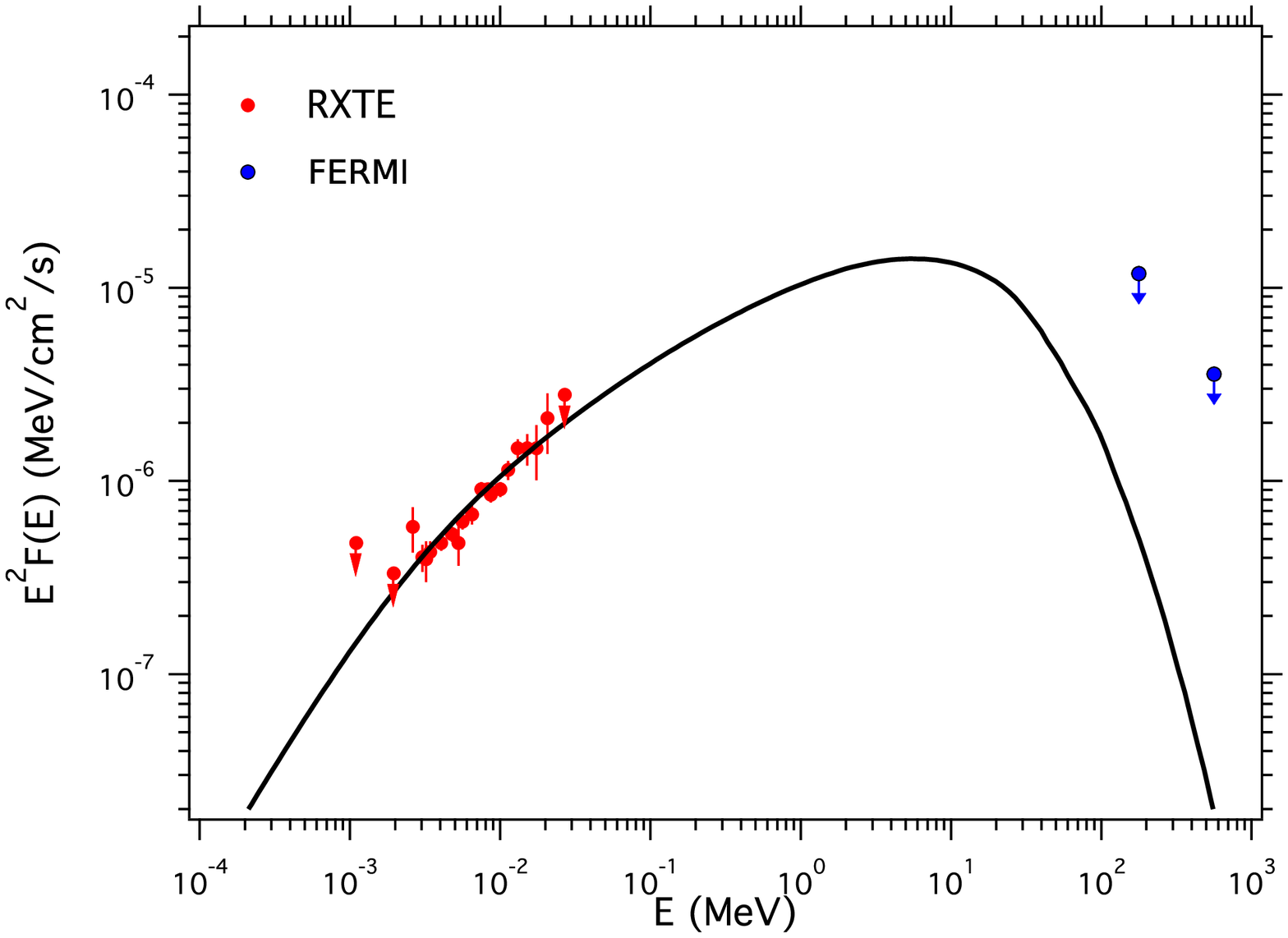}
\includegraphics[height=7cm,width=7cm]{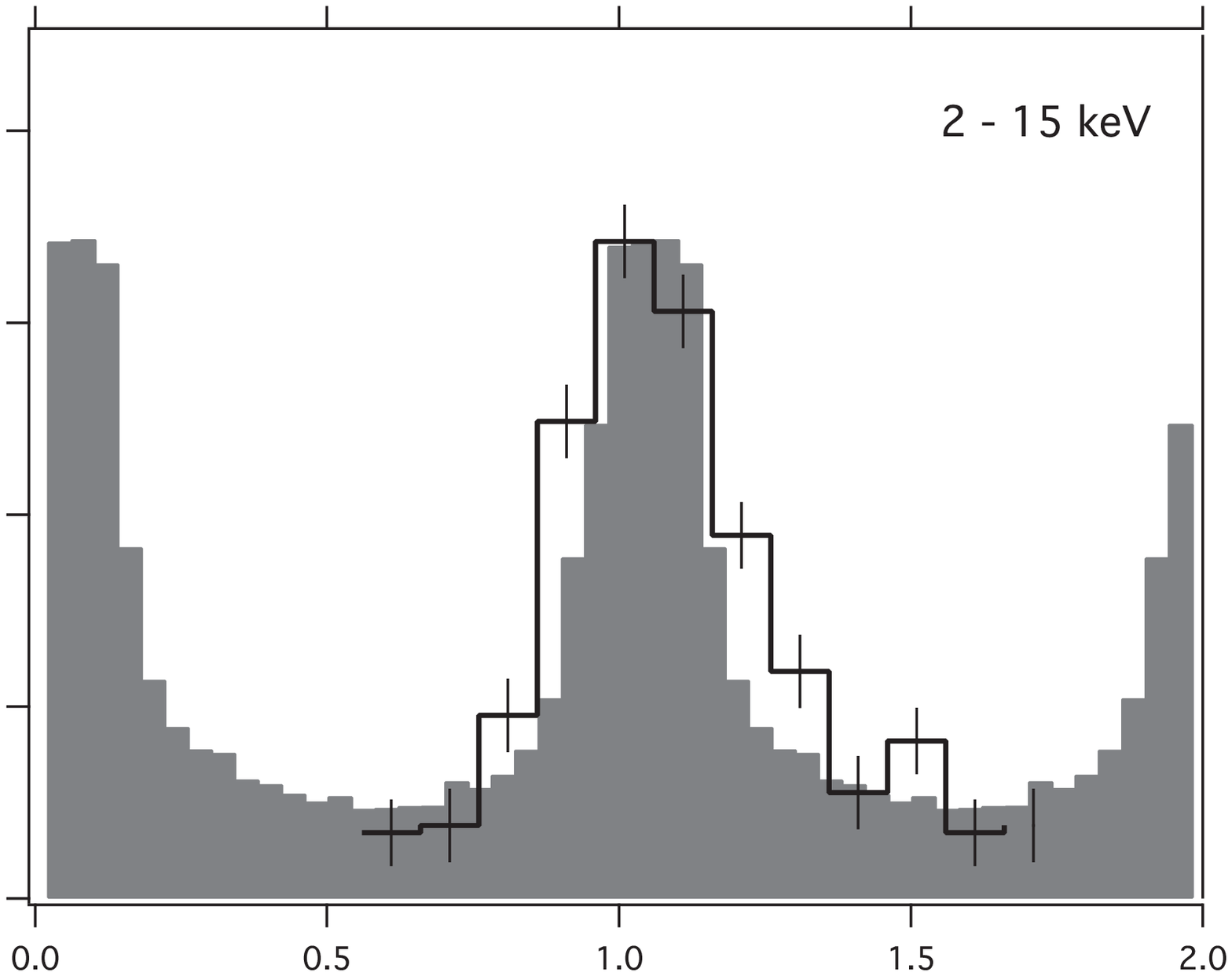}
\caption{Same as with Figure~\ref{1617}, but for PSR J1930+1852. 
The model calculation is result for $\alpha=20$degree and $\beta=41$degree, 
respectively. The data were taken  from Kuiper and Hermsen (2014) for observed 
X-ray flux and from Lu et al. (2007) for the X-ray light curve.}
\label{1930}
\end{center}
\end{figure}
The 136ms radio pulsar PSR J1930+1852 was found 
in SNR~G54.1+0.3 by Camilo et al. (2002).  Both the X-ray and radio pule 
profiles of PSR J1930+1852 show a single peak, but absolute phase 
difference between the X-ray and radio peaks has not been known. 
 The timing analyses show that the 
spin down age of this pulsar is 2.9kyr, and the surface dipole 
magnetic field is $B_s\sim 1\times 10^{13}$G. Lu et al. (2002) suggested that 
the distance of SNR G54.1+0.3 is about $d\sim 5$kpc by measuring the X-ray 
absorption column density. PSR J1930+1852 is surrounded by a PWN, which 
has clear torus and jet structure. The ratio between the observed semi-major 
 and semi-minor axes of the torus of PWN suggests that the Earth viewing angle 
inferred from the geometrical model is $\beta\sim 41$degree (Lu et al. 2002), 
which is  used in the calculation.  Figure~\ref{1930} compares
the calculated spectrum (left panel, solid line) and 
X-ray light curve (right panel, grey histogram) with the observations. 
We assumed the inclination angle of $\alpha=20$degree and the Earth viewing 
angle measured from the rotation axis of $\beta=41$degree. 

\section {Discussion and summary}
\label{summary}

The present model suggests that the 
observed X-ray/soft gamma-rays are the synchrotron emissions
from the high magnetic field region near the stellar surface, and
 the difference between the magnetic inclination
angle and the Earth viewing
angle is small, say $|\alpha-\beta|\le 30$~degree. To avoid the GeV emissions
from the outer gap, furthermore, the inclination angle is required to be
small, say $\alpha\le 30$~degree. We emphasize that 
the GeV-quit soft gamma-ray pulsars  actually emit outgoing  GeV gamma-rays
 from the outer gap, which make gamma-ray spectra of $Fermi$-LAT pulsars, 
but our line of sight is out of emission cone
 due to the smaller magnetic inclination and a smaller Earth viewing angle.
 For the Earth viewing angle 
is  $\beta\sim 40-50$~degree, the synchrotron emissions from
 the incoming pairs and the GeV emissions of 
out going particles can be observed, as we described in section~\ref{model}.
 This may be the case for GeV-loud soft gamma-ray pulsars PSRs J0205+6449
 and J2229+6114,  which show a very soft GeV spectra and
the smaller ratio of the GeV fluxes and X-ray fluxes comparing with
 the typical gamma-ray pulsars (e.g. Vela pulsar, 
Kuiper and Hermsen 2013, 2014). For the viewing angle
of $\beta\sim 70-90$~degree, the outward GeV emissions makes a spectral peak
in $\nu F_{\nu}$ at several GeV.   Within the framework of our scenario,
therefore, the viewing geometry is crucial 
factor to discriminate between the normal gamma-ray pulsars 
and soft gamma-ray pulsars, and the GeV-quiet SGPSR is peculiar case 
of the viewing geometry.

The GeV-quiet SGPSRs are relatively young 
and have higher-spin down powers compared with $Fermi$-LAT pulsars, 
as Figure~\ref{sgr} shows; the typical  characteristic age 
and the spin down power of the GeV-quiet SGPSRs are $\tau_s\sim 10^{3-4}$yr
 and $L_{sd}\sim 0.5-1\times 10^{37}{\rm erg~s^{-1}}$. 
With the current study, it is not obvious the reason why  GeV-quiet SGPSRs with
 the characteristic age of $\tau_s>10^{4}$yr and $L_{sd}<5\times 10^{36}
{\rm erg s^{-1}}$ have not yet found, while many typical gamma-ray pulsars 
with those spin down parameters have been found by the $Fermi$.  However, 
we expect that with the appropriate viewing geometry, 
 the pulsars with higher-surface magnetic field and/or high-spin down 
pulsars  are preferentially detected as the GeV-quiet SGPSRs.
 For the future study, therefore, we will study the 
evolution of spectrum 
in X-ray/gamma-ray bands with the viewing geometry, spin down parameters etc.
 and will discuss the population of the GeV-quiet soft gamma-ray pulsar, 
GeV-loud soft gamma-ray pulsars and typical gamma-ray pulsars.
 The linking among these three group of the gamma-ray
 pulsars with provide us a comprehensive picture of the high-energy  pulsars.

We thank Drs L. Kuiper and W.Hermsen for useful discussion and 
for providing us the X-ray/soft gamma-ray data. We express our appreciation 
to an anonymous referee for useful comments This work 
is partially supported by a 2014 GRF grant of 
Hong Kong Government under HKU 17300814P and Seed Funding Programme for Basic 
Research under HKU 2012-7159004 and 201310159026.

\end{document}